\documentclass[a4paper]{article}

\usepackage{INTERSPEECH2021}

\usepackage{amsfonts,amssymb} 

\usepackage{bbding}

\usepackage{caption} 
\usepackage{tabularx}
\usepackage{float}
\usepackage{array}  
\usepackage{booktabs} 
\usepackage{multirow}
\usepackage{makecell}
\usepackage{color}

\title{Attention-based cross-modal fusion for audio-visual voice activity detection in musical video streams}

\name{Yuanbo Hou$^1{^*}$, Zhesong Yu$^2$, Xia Liang$^2$, Xingjian Du$^2$, Bilei Zhu$^2$, Zejun Ma$^2$, Dick Botteldooren$^1$\thanks{$^*$ Work performed as an intern at Bytedance AI Lab.}}

\address{
  $^1$Ghent University, Belgium\\
  $^2$ByteDance AI Lab, China}
\email{\{yuanbo.hou, Dick.Botteldooren\}@UGent.be \\ \{yuzhesong, liangxia.21, duxingjian.real, zhubilei, mazejun\}@bytedance.com}

\begin{document}

\maketitle
\begin{abstract}
Many previous audio-visual voice-related works focus on speech, ignoring the singing voice in the growing number of musical video streams on the Internet. For processing diverse musical video data, voice activity detection is a necessary step. This paper attempts to detect the speech and singing voices of target performers in musical video streams using audio-visual information. To integrate information of audio and visual modalities, a multi-branch network is proposed to learn audio and image representations, and the representations are fused by attention based on semantic similarity to shape the acoustic representations through the probability of anchor vocalization. Experiments show the proposed audio-visual multi-branch network far outperforms the audio-only model in challenging acoustic environments, indicating the cross-modal information fusion based on semantic correlation is sensible and successful.
\end{abstract}

\section{Introduction}

With the popularity of musical videos on social platforms, a wide variety of musical videos have been uploaded to the Internet. To recognize speech and singing voices in these videos, voice activity detection (VAD) is a necessary preprocessing to identify the start and end time of human voice activities. VAD has attracted many interests due to its wide applications such as speech \cite{speechcoding, speechrecognition} and music information processing \cite{Hou2020}.

In scenes of musical video streams, usually one or more anchors (performers) sing or talk to the audience in front of the camera, while music is being played in the background, containing voices of other people and accompaniments. Besides, there may be other sounds such as applause, cheers, and screams from audiences. This paper aims to detect the singing voice and speech of an anchor (the target performer) in musical videos, which have a challenging acoustic environment like a cocktail party. In such noisy environments, audio-only VAD methods \cite{sohn1999statistical, hmm, ishizuka2010noise, lee2018revisiting} are difficult to work accurately and effectively because the sound of various musical instruments, cheers and applause from audiences, and other non-target sound signals, will interfere with the audio-only VAD. Besides, voices from other non-target people in the background music can be mistaken as active voices. It is difficult to distinguish the target voice from various non-target sounds using only audio information, and it is more difficult to further identify the target singing voice and speech when they are embedded in transient interferences and highly non-stationary noises \cite{dov2015audio} from the background. That is, audio-only VAD methods do not work well in musical videos.

The visual information from video is more robust than the audio information from noisy acoustic environments. The facial information of an anchor can directly reflect whether the anchor is vocalizing. So, audio-visual VAD (AVVAD) is introduced to utilize the visual modality to make up for deficiencies of audio-only VAD in complex acoustic environments. Visual features are used to identify natures of lips in AVVAD \cite{liu2006audio} for speech processing. By analyzing lip shapes during speech and non-speech, an appropriate visual parameter \cite{sodoyer2006analysis} is used for detecting sections of voice activity in speech embedded in non-stationary noise. Even in clean acoustic conditions using visual channels in addition to speech results in significantly improved classification performance \cite{mroueh2015deep}. However, the above audio-visual works mainly focus on speech but not on general sounds such as music and singing voice. This paper aims to detect the anchor's speech and singing voice in musical video streams, which is more challenging because there are not only a lot of speech-like noises but also other people's voices in audio streams.

To ignore the interference of acoustic noise and pay attention to detect voices of the anchor (target performer) in musical videos, this paper uses visual information that is not affected by acoustic noises to assist the model to more accurately judge voices source. Therefore, how to fuse information between two modalities to achieve a better combination effect is the core challenge of this work. Audio-visual integration strategies in previous works can be divided into three categories: feature fusion (\textsl{FF}) \cite{huang2013audio}, decision fusion (\textsl{DF}) \cite{teissier1999comparing}, intermediate fusion (\textsl{IF}) \cite{receveur2016turbo}. \textsl{FF} is simple splicing of audio and visual features to form a new feature set and modeling it. Based on the modeling of audio and image stream respectively, \textsl{DF} controls the final decision result by stream weight. \textsl{IF} attempts to model the fusion of intermediate representations of audio and visual features. Compared with \textsl{IF}, \textsl{DF} cannot take advantage of the temporal and semantic correlation between audio and visual features. To exploit the correlation between audio and visual features in musical videos, this paper uses \textsl{IF} to integrate audio and visual vectors and make comprehensive decisions, that is, high-level representations of acoustic and image features are fused with the attention mechanism based on the semantic similarity.

\label{ssec:figure-f}
\begin{figure*}[ht]
	\vspace{-0cm}  
	\setlength{\abovecaptionskip}{0.1cm}   
	\setlength{\belowcaptionskip}{-0.6cm}   
	\centerline{\includegraphics[width = 0.8 \textwidth]{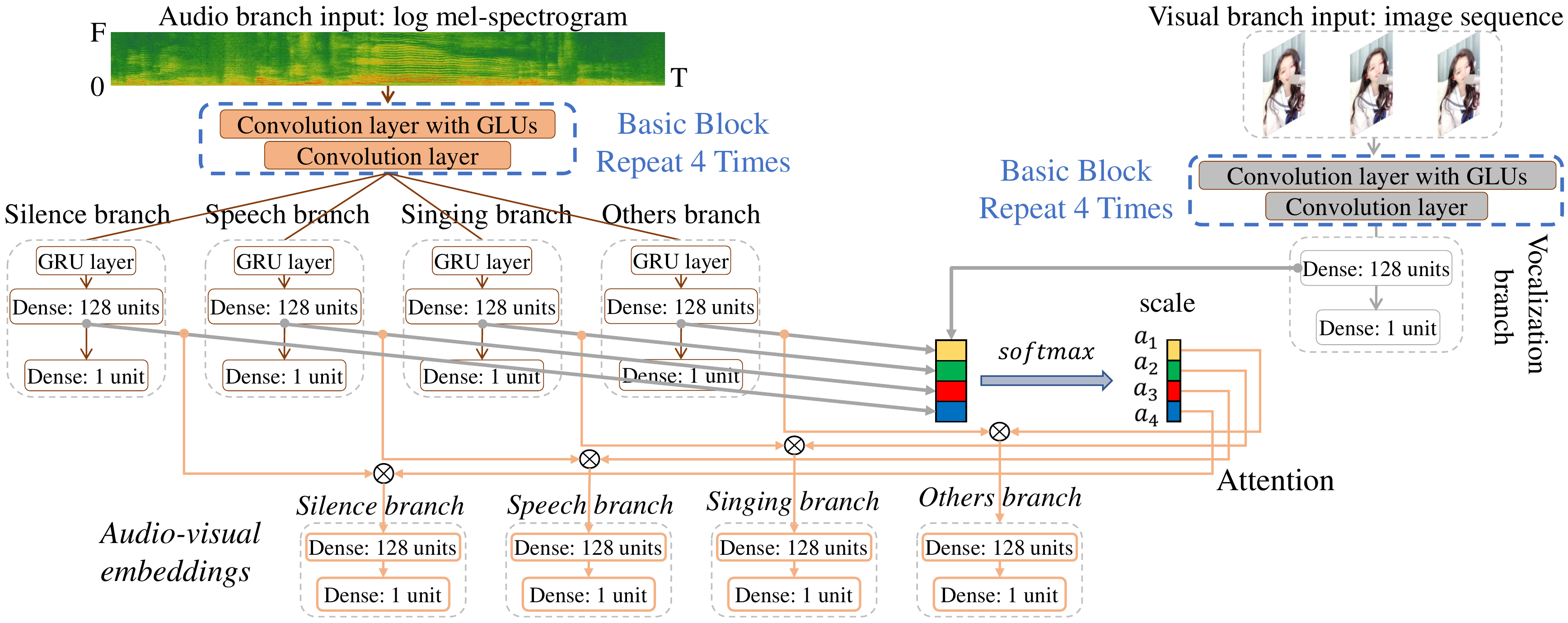}}
	\caption{The proposed attention-based AVVAD (ATT-AVVAD) framework.}
	\label{VAD}
\end{figure*}

In this paper, the correlation in the semantic space between the voice representations and the anchor vocalization representations is used to determine whether the voice comes from the anchor. The voice representations are adjusted according to the corresponding correlation coefficient based on attention. This paper attempts to explore the possibility of cross-modal fusion based on semantic similarity between different modalities to help the model actively learn how to fuse cross-modal information, let the model decide “how” to combine given multi-modal information most optimally, rather than based on artificial rules.

The main contributions of this paper are: 1) a multi-branch network is proposed for AVVAD to learn the high-level representations of different target events, and fuse the cross-modal information based on the semantic correlation by attention; 2) the possibility of detecting both the speech and singing voice of the target performer in challenging noisy acoustic environments is explored; 3) the intermediate representations of the proposed AVVAD model are visually analyzed to further investigate the performance of the model. This paper is organized as follows, Section 2 shows the attention-based audio-visual framework. Section 3 describes the dataset, baseline, experimental setup, and analyzes the results. Section 4 gives conclusions.

\vspace{-0.2cm}
\section{Multi-branch ATT-AVVAD framework}
The proposed attention-based AVVAD (ATT-AVVAD) framework in Figure~\ref{VAD} consists of the audio-based module, image-based module, and attention-based fusion module. The audio-based module produces acoustic representation vectors for four target audio events: Speech of the anchor, Singing voice of the anchor, Silence, and Others. The image-based module aims to obtain the possibility of anchor vocalization based on facial parameters. Finally, an attention-based module fuses audio-visual information to comprehensively consider the bi-modal information to make final decisions at the audio-visual level.

\vspace{-0.2cm}
\subsection{The audio-based module (audio branch)}

The goal of the audio-based module is to predict the probability of four target event classes (Silence, Speech, Singing, Others) at the audio level, wherein Singing and Speech only refer to the singing voice and speech of the anchor, rather than those from other people. Different from typical sound event detection (SED) \cite{xu2018large} models, the audio branch in Figure~\ref{VAD} attempts to generate high-level core acoustic representations for each target event class at the end of each output. In Figure~\ref{VAD}, a multi-output convolutional recurrent neural network (CRNN) is used to extract the core representations of different event classes. The log mel-spectrogram \cite{logmel} is extracted from the audio clip and input into the network. Then there are four blocks and each block contains gated linear units (GLUs) \cite{GLU}, a convolutional layer, a batch normalization layer \cite{batch}, and a ReLU \cite{ReLU}. GLUs can be used to effectively learn local shift-invariant patterns and acoustic representations of target events from the spectrogram \cite{hou2019sound}. 

To obtain a separate acoustic representation vector for each target event, after the blocks there are four independent embedding layers to extract core representation. Each embedding layer includes a GRU layer to capture temporal information, followed by two fully connected layers. The output of the first fully connected layer is regarded as the core representation vector of the corresponding event, and will be used to combine with the visual embedding vector. The second fully connected layer with one unit is a binary classification with sigmoid \cite{sigmoid} to predict the probability of the corresponding event in the current audio branch input. Please visit the source code on our homepage ({https://github.com/Yuanbo2020/Attention-based-AV-VAD}) for the specific parameters and real video detection demos.


\vspace{-0.17cm}
\subsection{The image-based module (visual branch)}

In musical videos, there are both a lot of speech-like instrumental accompaniment sounds and other people’s voices. These interferences result in the poor performance of relying on the audio signal to detect the anchor's speech and singing voice because it is difficult to determine the source of voices with audio alone. Hence, it is necessary to combine the anchor's visual information, such as the change of eye and lip contours, to distinguish whether voices are from the anchor or other people.

The visual branch aims to obtain the core representation vector of the anchor vocalization and to assist in judging whether voices are from the anchor. To obtain the probability of the anchor vocalization, a fixed-length image sequence corresponding to the input time of the audio branch is input to the network. The visual branch and audio branch have a similar structure with different parameters. Experiments \cite{singhal2018comparing} show convolutional layers are more effective than GRU layers in image feature extraction, so there are no GRU layers in the visual branch. Similar to the structure of the audio branch, there are two fully connected layers after four blocks in the visual branch. The output of the dense layer with 128 units is regarded as the core representation vector of the anchor vocalization, the final classification layer with one unit is a binary classification with sigmoid to predict the probability of the anchor vocalization.

\vspace{-0.17cm}
\subsection{Attention-based fusion module}

In videos, visual events usually occur together with acoustic events and they are coordinated. The facial information can reflect whether the anchor is vocalizing or not. To explore the correlation between the anchor’s voice activity and face parameters, this paper attempts to fuse the audio and visual high-level representation vectors based on the semantic similarity by attention mechanism to make comprehensive decisions for detecting the four target event classes at the audio-visual level.

In Figure~\ref{VAD}, the subbranches in the audio branch output the high-level representation vector of Silence, Speech, Singing, and Others, respectively. The visual branch outputs the representation vector to represent the anchor vocalization. By the attention module, the correlation between acoustic representation vectors and visual vocalization vector in the semantic space is calculated to strengthen audio-level representations, then the network can pay more attention to sound events related to the anchor. Given \{$\boldsymbol{q_1}$, $\boldsymbol{q_2}$, $\boldsymbol{q_3}$, $\boldsymbol{q_4}$\}${\in\mathbb{R}^{128\times1}}$ denote the acoustic embedding vector of Silence, Speech, Singing and Others in the audio branch, respectively. $\boldsymbol{Q}=[\boldsymbol{q_1}, \boldsymbol{q_2}, \boldsymbol{q_3}, \boldsymbol{q_4}]$. $\boldsymbol{K} {\in\mathbb{R}^{128\times1}}$ denotes the visual vocalization vector. The attention ($ATT$) \cite{attention} can be defined as:
\begin{equation}
\setlength{\abovedisplayskip}{3pt}
\setlength{\belowdisplayskip}{3pt}
ATT = Softmax({Q^TK}/{\sqrt{d_k}})
\end{equation}
where $d_k$ is the dimension of $\boldsymbol{K}$. $ATT$ is a ${4\times1}$ vector containing the scaling factors to be applied to the representation vector of the corresponding acoustic event. After multiplying the acoustic event vector with the corresponding attention scale factor, the audio-visual event vector is obtained.

\begin{figure}[t]
	\vspace{-0cm}  
	\setlength{\abovecaptionskip}{0cm}   
	\setlength{\belowcaptionskip}{-0.5cm}   
	\centerline{\includegraphics[width = 0.44  \textwidth]{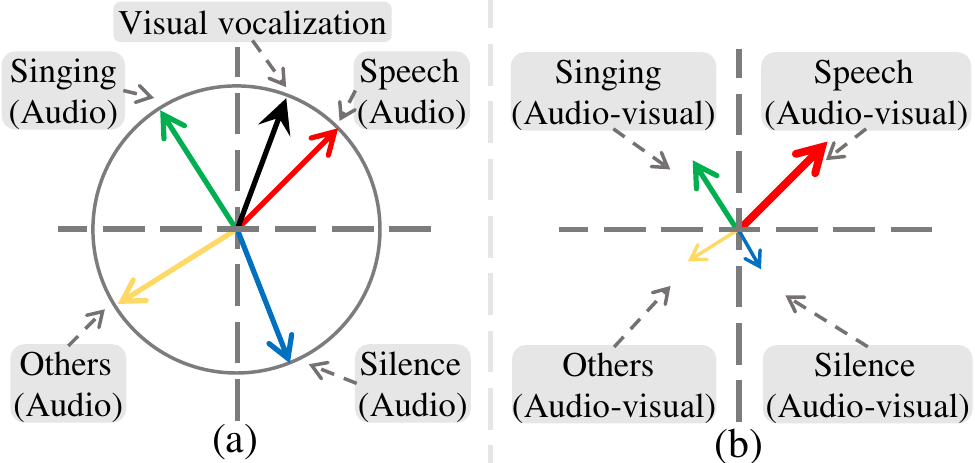}}
	\caption{The main idea of attention-based fusion. Subgraph (a) shows the audio and visual representation vectors in the same semantic space. Subgraph (b) is the final audio-visual representation vector of target events after the attention-based fusion.}
	\label{attention}
\end{figure}

Figure~\ref{attention} shows the core idea of the attention-based fusion, when the audio branch detects the speech or singing voice, and the visual branch predicts the anchor is vocalizing at this time, the corresponding acoustic event vector will be given more attention and transmitted to the audio-visual module. With the scaling effect of the attention factor, we try to train the model to find such a relationship: when the audio branch detects speech or singing voice, and the visual branch indicates the anchor is vocalizing, then representations of the speech or singing voice will be relatively enhanced, that is, representations of the audio branch is confirmed in the audio-visual module. Representations given by the audio branch are corrected, and errors of the audio-visual module are reduced. That is, only when representations of the audio and visual branch are consistent in the semantic space, the corresponding acoustic representations will be relatively enhanced, while the importance of other acoustic event representations will be relatively weakened.

To consider the information of audio, visual, and audio-visual modality at the same time in the training phase, the losses of different modules are calculated together. The final loss function of the ATT-AVVAD model is:
\begin{equation}
	\setlength{\abovedisplayskip}{2.5pt}
	\setlength{\belowdisplayskip}{2.5pt}
	\begin{split}
	L=&\lambda_1L_{a-sil}+\lambda_2L_{a-spe}+\lambda_3L_{a-sin}+\lambda_4L_{a-oth}\\
	&+\lambda_5L_{v-voc} + \lambda_6L_{av-sil} + \\
	&\lambda_7L_{av-spe} + \lambda_8L_{av-sin} + \lambda_{9}L_{av-oth} \\
	\end{split}
\end{equation}
where $L_a$, $L_v$ and $L_{av}$ denote the loss of audio branch, visual branch and audio-visual module; $sil$, $spe$, $sin$ and $oth$ denote silence, speech, singing and others; $voc$ denotes vocalizing. $\lambda_i$ is the scale factor of each loss function, the size of $\lambda_i$ determines the importance of each loss function in training.

\vspace{-0.15cm}
\section{Experiments and results}

\vspace{-0.05cm}
\subsection{Dataset, Baseline, and Evaluation metrics}
To train the ATT-AVVAD model to detect both target speech and singing voice in challenging acoustic environments, a 500-minute video dataset with frame-level labels is used. The duration of the dataset for training, validation, and testing is 360 mins, 40 mins, and 100 mins, respectively. To prevent the model bias caused by the unbalanced number of male and female anchors in the training phase, the total duration of live broadcasts of male and female anchors is close in the dataset.

\label{ssec:attention}

In training, log mel-bank energy with 64 banks \cite{mel} of the input audio stream is used in the audio branch, which is extracted by STFT with Hamming window length of 44 ms and overlap of 50\% between the window. To comprehensively consider the contextual information, the input of audio branch is a moving feature block whose time length is consistent with that represented by the image sequence of the visual branch. ATT-AVVAD model is expected to be used online, a short latency is required. So a lightweight face detection algorithm \cite{facenet} is used to pre-mark the face in visual inputs, to reduce the computational burden of the model and help it to find focal areas faster and more accurately. The output of each branch in ATT-AVVAD is binary classification, hence Adam optimizer \cite{adam} with a learning rate of 0.001 is used to minimize the binary cross-entropy. Dropout \cite{dropout} is used to prevent overfitting.

To compare the performance of our proposed method comprehensively, two baseline systems are considered. A common and typical multi-modal recurrent neural model \cite{tao2019end} with two branches similar to the proposed ATT-AVVAD is used as the audio-visual baseline (\textsl{Base-AV}). A compound convolutional recurrent neural network \cite{Hou2020} trained by transfer learning is used as the audio-only baseline (\textsl{Base-A}) to compare the performance of the ATT-AVVAD from more perspectives.

For evaluation metrics, event-based precision (\textsl{P}), recall (\textsl{R}), \textsl{F-score} and Error rate (\textsl{ER}) \cite{metrics} are used to analyze the performance of the model. Compared with segment-based metrics used in previous studies \cite{Mesaros2019_TASLP}, event-based metrics are more rigorous and accurate to measure the location of events. Higher \textsl{P}, \textsl{R}, \textsl{F} and lower \textsl{ER} indicate a better performance.

\vspace{-0.2cm}
\subsection{Results and Analysis}
This section tries to analyze the performance of the proposed method based on the following \textbf{R}esearch \textbf{Q}uestions (\textbf{RQ}):

\noindent
\textbf{• RQ1}: $\lambda_i$ in the final loss function determines the importance of each loss function in training \cite{multi-loss}. What effect do different values of $\lambda_i$ have on the performance of the model?

\vspace{-0.03cm}
\begin{table}[b]\footnotesize
	\setlength{\abovecaptionskip}{0cm}   
	\setlength{\belowcaptionskip}{-0.55cm}   
	\renewcommand\tabcolsep{1.5pt} 
	\centering
	\caption{Results of different values of $\lambda_i$ on the test dataset.}
	\begin{tabular}{p{2.2cm}<{\centering}p{1.1cm}<{\centering}p{2.2cm}<{\centering}p{1.8cm}<{\centering}} 
		\toprule[1pt] 
		\specialrule{0em}{0.1pt}{0.1pt}
		{\{$\lambda_1$, $\lambda_2$, $\lambda_3$, $\lambda_4$\}} & {\{$\lambda_5$\}} & {\{$\lambda_6$, $\lambda_7$, $\lambda_8$, $\lambda_9$\}} & \textsl{F-score (\%)}\\
		\midrule[1pt]  
		
		\specialrule{0em}{0.2pt}{0.2pt}
		\textsl{\{1, 1, 1, 1\}} & \{1\} & \{0.5, 0.5, 0.5, 0.5\} & 76.38\\
		
		\specialrule{0em}{0.2pt}{0.2pt}
		\textsl{\{1, 1, 1, 1\}} & \{0.5\} & \{1, 1, 1, 1\} & 77.01\\
		
		\specialrule{0em}{0.1pt}{0.1pt}
		\textsl{\{0.5, 0.5, 0.5, 0.5\}} & \{1\} & \{1, 1, 1, 1\} & 76.73\\
		
		\specialrule{0em}{0.1pt}{0.1pt}
		\textsl{\{1, 1, 1, 1\}} & \{1\} & \{1, 1, 1, 1\} & 77.81\\
		
		\specialrule{0em}{0.2pt}{0.2pt}
		\textsl{\{0.5, 0.5, 0.5, 0.5\}} & \{0.5\} & \{1, 1, 1, 1\} & 77.90\\
		
		\specialrule{0em}{0.2pt}{0.2pt}
		\textsl{\{0.5, 0.5, 0.5, 0.5\}} & \{0.5\} & \{0.5, 0.5, 0.5, 0.5\} & 77.12\\
		
		\bottomrule[1pt]
	\end{tabular}
	\label{tab:loss}
\end{table}

The four classes of target events in our task are equally important, so $\lambda_i$ related to the event sub-branch in the audio branch are the same in Table \ref{tab:loss}. The same goes for audio-visual branch. In Table \ref{tab:loss}, given $\lambda_1$ is 1 and $\lambda_5$ is 0.5, it is equivalent to attaching importance to the audio silence branch and reducing the importance of the visual vocalization branch. Different values of $\lambda_i$ represent the difference in importance between audio
information, visual information, and audio-visual information. Table \ref{tab:loss} shows that for the VAD task, the joint audio-visual information is more important than the audio information, and the audio information is more important than the visual information.

\noindent
\textbf{• RQ2}:
Does the proposed ATT-AVVAD in this paper perform better than the audio-visual baseline \textsl{Base-AV}? Are the detection results of ATT-AVVAD better than audio-only VAD baseline \textsl{Base-A}, and how much improvement is there?

To compare the performance of the proposed method, Table \ref{tab:rule} shows the detailed results of the proposed ATT-AVVAD, \textsl{Base-AV} and \textsl{Base-A}. The last classification layer of \textsl{Base-AV} uses the Softmax activation function to make decisions, which means that each event class in the classification layer is equally important. As mentioned before, $\lambda_i$ is the scale factor of each loss function. For a fair comparison between the proposed ATT-AVVAD model and audio-visual baseline \textsl{Base-AV}, all $\lambda_i$ in $L$ are equal to 1 in the following results.

\vspace{-0.03cm}
\begin{table}[b]\footnotesize
	\setlength{\abovecaptionskip}{0cm}   
	\setlength{\belowcaptionskip}{-0.55cm}   
	\renewcommand\tabcolsep{1.5pt} 
	\centering
	\caption{Event-based evaluation of detection results.}
	\begin{tabular}{p{1.9cm}<{\centering}p{1.2cm}<{\centering}p{1.2cm}<{\centering}p{1.2cm}<{\centering}p{1.9cm}<{\centering}} 
		\toprule[1pt] 
		\specialrule{0em}{0.1pt}{0.1pt}
		& \textsl{ER} & \textsl{P (\%)} & \textsl{R (\%)} & \textsl{F-score (\%)}\\
		\midrule[1pt]  
		\specialrule{0em}{0.1pt}{0.1pt}
		\textsl{Base-A} & 0.86 & 47.93 & 35.72 & 40.93\\
		\specialrule{0em}{0.1pt}{0.1pt}
		\textsl{Base-AV} & 0.72 & 66.17 & 53.41 & 59.11\\
		\specialrule{0em}{0.2pt}{0.2pt}
		\textsl{ATT-AVVAD} & \textbf{0.39} & \textbf{85.24} & \textbf{71.57} & \textbf{77.81}\\
		\specialrule{0em}{0.2pt}{0.2pt}
		\bottomrule[1pt]
		\specialrule{0em}{0pt}{0pt}
	\end{tabular}
	\label{tab:rule}
\end{table}

The results in Table \ref{tab:rule} show that: 1) in challenging noisy environments like musical video, the performance of anchor's voices detection based on audio-visual information (\textsl{Base-AV} and \textsl{ATT-AVVAD}) far outperforms that of the audio only approach (\textsl{Base-A}). Compared with the audio-only \textsl{Base-A} with an \textsl{ER} of 0.86 and \textsl{F-score} of 40.93\%, the \textsl{ER} and \textsl{F-score} of the ATT-AVVAD in this paper are 0.39 and 77.81\%. 2) Even though both are based on audio-visual information, the cross-modal learning method based on attention fusion proposed in this paper has a lower \textsl{ER} and more accurate detection results than the typical audio-visual model \textsl{Base-AV}, which means the bi-modal framework proposed in this paper is effective, and the attention-based fusion mechanism is helpful.

\label{ssec:tsne}
\begin{figure}[t]
	\vspace{0cm}  
	\setlength{\abovecaptionskip}{0.2cm}   
	\setlength{\belowcaptionskip}{-0.4cm}   
	\centerline{\includegraphics[width = 0.5  \textwidth]{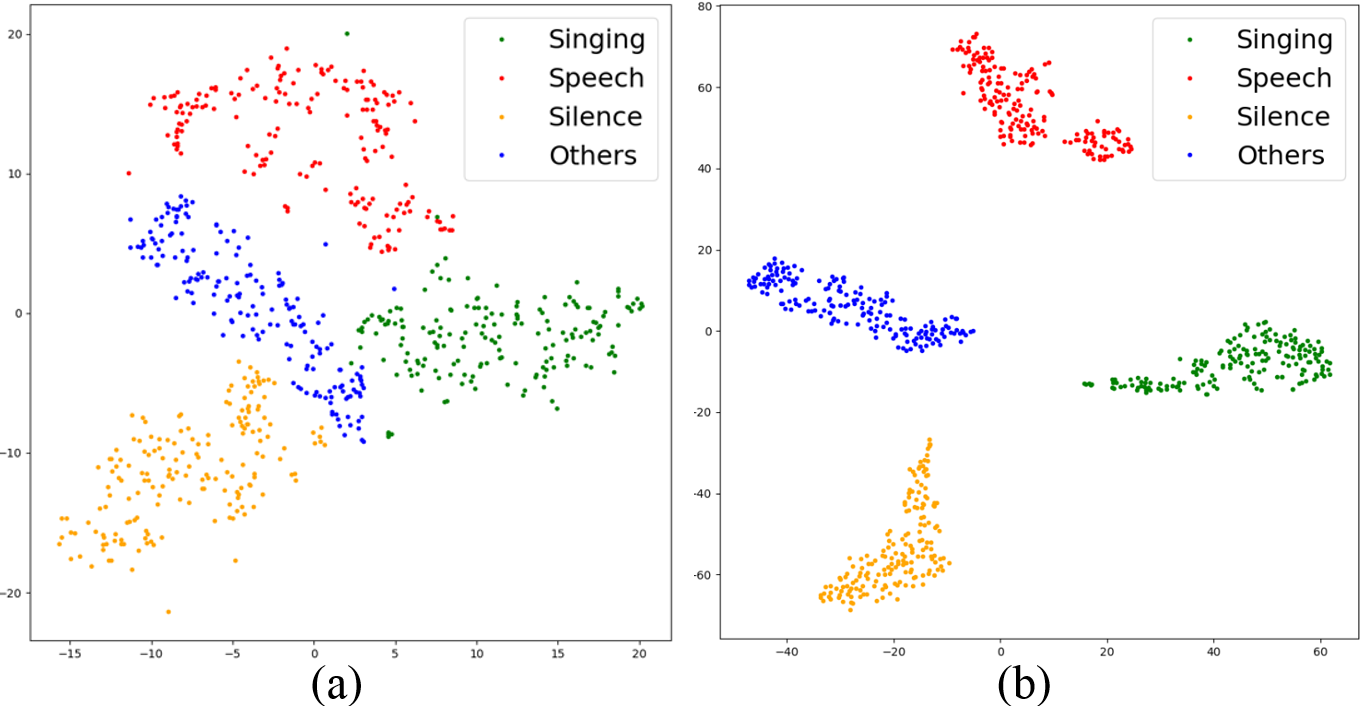}}
	\caption{Visualization of core representation vectors distribution from a test sample using t-SNE \cite{tsne}. The vectors in subgraph (a) are from the audio branch, vectors in subgraph (b) are from audio-visual modules after attention-based fusion.}
	\label{tsne}
\end{figure}

\noindent
\textbf{• RQ3}: What changes have taken place in the model learning before and after the attention-based fusion?


\label{ssec:tsne2}
\begin{figure}[!t]
	\setlength{\abovecaptionskip}{0.2cm}  
	\setlength{\belowcaptionskip}{-0.4cm}   
	\centerline{\includegraphics[width = 0.25 \textwidth]{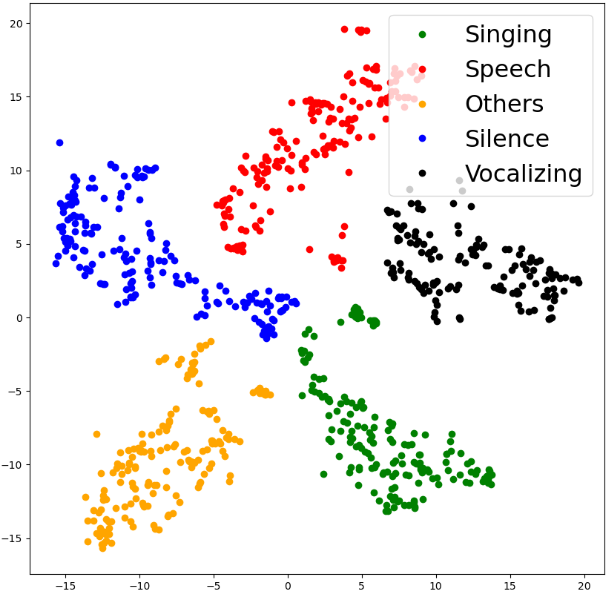}}
	\caption{Visualization of acoustic representation vectors and visual vocalization vector distribution from a test sample using t-SNE. The vector (black dots) representing the vocalizing of the anchor is distributed on the side representing the voices of the anchor (green dots for singing, red dots for speech).}
	\label{tsne2}
\end{figure}

To gain deeper insights into the effect of attention-based fusion on the model, the distribution of the core representation vector representing four target event classes before and after the fusion is visualized. The representation vectors of target events in the audio branch can reflect the decision tendency of the model before the fusion, and the corresponding audio-visual representation vectors can reflect modified results of the model after the attention fusion. As shown in Figure~\ref{tsne}, before fusion, the model can roughly divide different target classes in the audio branch, but the classification boundary interval between each class is not obvious, and each cluster is not compact. After focusing on the visual vector based on attention, there is a clear classification boundary between different classes, and each class is more tightly clustered. This indicates the proposed attention-based fusion does play a regulating role and makes the final joint classification in the audio-visual module easier. Based on attention fusion, the distribution of acoustic core representation vectors and visual vocalization vector in semantic space tends to be aligned as shown in Figure~\ref{tsne2}. The visual vocalization vector is distributed on the side of the vectors of speech and singing voice, and is away from the event vectors without the anchor's voices, which means the semantics of audio and visual vectors are consistent and the cross-modal information fusion based on the correlation between acoustic embeddings and visual vocalization vectors is reasonable.

The bi-modal ATT-AVVAD is effective, but how big is the effect of different modal branches? The ablation studies \cite{ablation} in Table \ref{tab:ablation} show the results of ATT-AVVAD after removing certain structures. The detection result based on pure visual information is the worst, perhaps because the opening and closing of the mouth of the anchor are binary, it is difficult to use the binary information to detect four types of target events. Compared with the audio-only detection results, the bi-modal detection results have been improved, indicating that it is useful to use visual information to correct audio information to assist judgment.

\vspace{-0.03cm}
\begin{table}[b]\footnotesize
	\setlength{\abovecaptionskip}{0cm}   
	\setlength{\belowcaptionskip}{-0.55cm}  
	\renewcommand\tabcolsep{1pt} 
	\centering
	\caption{Ablation experiments of the ATT-AVVAD model.}
	\begin{tabular}{p{1.9cm}<{\centering}p{1.9cm}<{\centering}p{1.9cm}<{\centering}p{1.7cm}<{\centering}}
	
		\toprule[1pt] 
		\specialrule{0em}{0.1pt}{0.1pt}
		{Audio module}& \textsl{Visual module} & \textsl{Fusion module} & \textsl{F-score (\%)}\\
		\midrule[1pt]  
		\specialrule{0em}{0.pt}{0.pt}
		\CheckmarkBold & \XSolidBrush & \XSolidBrush & 68.59\\
		
		\specialrule{0em}{0.pt}{0.pt}
		\XSolidBrush & \CheckmarkBold & \XSolidBrush & 33.92\\
		
		\specialrule{0em}{0pt}{0pt}
		\CheckmarkBold & \CheckmarkBold & \CheckmarkBold & 77.81\\
		\bottomrule[1pt]
	\end{tabular}
	\label{tab:ablation}
\end{table}




\vspace{-0.1cm}
\section{Conclusion}
\label{sec:CONCLUSION}

To detect the speech and singing voice of the anchor in musical video streams, a multi-branch ATT-AVVAD framework is proposed with attention-based fusion by semantic similarity, which performs well in noisy environments. Experiments show that: 1) the performance of the audio-visual model far outperforms that of the audio-only model in challenging acoustic environments; 2) the multi-branch network that can produce core representation vectors for target events, and attention-based fusion are effective; 3) the cross-modal information fusion based on semantic similarity is successful.

\vspace{-0.1cm}
\section{Acknowledgments}
\label{sec:ACKNOWLEDGEMENTS}
This research received funding from the Flemish Government under the “Onderzoeksprogramma Artificiële Intelligentie (AI) Vlaanderen” programme.

\vfill\pagebreak
\bibliographystyle{IEEEtran}

\bibliography{template}

\end{document}